\documentclass[preprint,10pt,plainnat]{aastex6}
\usepackage{amsmath,amstext} 
\usepackage[figure,figure*]{hypcap} 
\usepackage{newtxmath}
\usepackage{natbib}

\shorttitle{Extremophile life-form survey on rocky exoplanets} 
\shortauthors{Madhu Kashyap Jagadeesh}

\begin{document}

\title{Extremophile life-form survey on rocky exoplanets}
\author{Madhu Kashyap Jagadeesh \altaffilmark{1}}
\email{kas7890.astro@gmail.com}
\altaffiltext{1}{Department of Physics, Jyoti Nivas College, Bengaluru-560095, Karnataka, India}

\begin{abstract}
Search for different life-forms elsewhere is the fascinating area of research in astrophysics and astrobiology. Nearly 3500 exoplanets are discovered according to NASA exoplanet archive database. Earth Similarity Index (ESI) is defined as the geometrical mean of radius, density, escape velocity and surface temperature, ranging from 0 (dissimilar to Earth) to 1(Earth). In this research, rocky exoplanets that are suitable for rock dependent extremophiles, such as: Chroococcidiopsis and Acarosporamto are chosen, which can potentially survive are considered. The Colonizing Similarity Index (CSI) is introduced and analysed for 1650 rocky exoplanets, CSI is basically representing Earth-like planets that are suitable for rocky extremophiles which can survive in extreme temperatures (i.e. as hot as desert and cold as frozen lands). In this work the in-habitable exoplanets are recognised even for these rocky extremophiles to not potentially survive by using the CSI metric tool.
\end{abstract}


\section{Introduction}
Extraterrestrial research in recent years is the Holy Grail of current sciences. The space missions like CoRoT and Kepler has directed a huge flow of data from exoplanetary observations, which is maintained by Planetary Habitability Laboratory- Exoplanet catalogue (PHL-EC, 2017). The PHL-EC data of planetary objects, such as: radius, density, escape velocity and surface temperature has been applied as the geometrical mean to a metric tool called ESI, which ranges from 0 (dissimilar to Earth) to 1 (similar to Earth) (Schulze-Makuch et al., 2011). This tool helps us to identify the potentially habitable planets (php’s) or Earth-like planets from the observed physical parameters of extra-solar systems. 
The exoplanetary composition is basically classified into rocky and gas giants. For planets to be rocky the mass  ranges from 0.1 to 10 Earth masses and radius is of the order 0.5 to 2 Earth radii (Safonova et al., 2016). Recently, Kashyap et al. (2017) have introduced a new technique to find the surface temperature of exoplanets and formulated the Mars Similarity Index (MSI) for search of extremophile life-forms in Mars-like conditions. In this current work, we focus on rocky exoplanets in Earth-like conditions, by varying the surface temperature parameter, which is suitable for the Extremophile life-forms to reproduce and grow. Hence the Colonizing Similarity Index (CSI) is formulated and analysed for 1650 rocky exoplanets. 
The structure of the paper is as follows, the extremophiles and ESI are discussed in section 2 and section 3 respectively. Section 4 illustrates the CSI analysis, followed by section 5 which contains the discussion and conclusion part of the work.
 
\section{Extremophiles}

“Extremophiles” are those microbes which love extreme environments such as, very high and low temperatures. Based on the observations from the evolution of earth, the limitations of survival characteristics are defined, and compared with conditions of exoplanets (Hegde et al., 2013). The targeted extremophiles in this paper are Chroococcidiopsis and Acarosporamto, whose growth depends on rocky surface (Cumbers and Rothschild, 2014; Nash et al., 2001).
Chroococcidiopsis is a micro-organism that is tested for survival capability in outer space. The different kinds of tests performed on Chroococcidiopsis are namely: low Earth orbit, impact event, planetary ejection, atmospheric re-entry, and simulated conditions. (Cockell et al., 2005; Billi, 2011; Baqué et al., 2013; Cockell et al., 2011).
In 2014 an Expose astrobiology lab was launched and mounted on the International space station (Gerda Horneck et al., 2013, Karen Olsson-Francis et al., 2013). According to Nash et al., 2001 and Onofri et al., 2015 Acarosporamto, a lichenized fungal genus survived in this Martian condition outside Earth.   

\section{Earth Similarity Index (ESI)}
The well-studied indexing introduced in 2011 by Schulze-Makuch et al., and recently Kashyap et al. 2017, has derived ESI from the threshold value (V) in the similarity scale for each quantity, which is mathematically expressed as, 

\begin{equation}
V = \left[1-\Big|\frac{x_0-x}{x_0+x}\Big| \right]^{w_x}\,.
\label{SIIeq1}
\end{equation}
where $x$ is the planetary property of the exoplanet, $W_X$ is the weight exponent and $x_0$ is the reference to Earth in ESI. Assuming the threshold value of V=0.8 (~80 \% similar to the reference) and defining the physical limits xa and xb as the variation with respect to x0 (ie, xa<x0< xb), the weight exponents are computed as lower wa and upper wb limits, which is expressed as:

\begin{equation}
w_a = \frac
{\ln{V}}
{\ln\left[1-\left|\frac{x_0-x_a}
{x_0+x_a}\right|\right]}
\,,\quad 
w_b = \frac
{\ln{V}}
{\ln\left[1-\left|\frac{x_b-x_0}{x_b+x_0}\right|\right]}
\,,\quad 
\label{eq:weight_exponent}
\end{equation}
The average weight can be written as,

 \begin{equation}
w_x=\sqrt{{w_a}\times {w_b}}\,.
\end{equation}

Hence the Earth Similarity Index can be obtained as:
\begin{equation}
ESI_x = {\left[1-\Big|
\frac{x-x_0}{x+x_0}\Big| \right]^{w_x}}\,,
\label{eq:esi}
\end{equation}
where, x is the physical parameter of the exoplanet, such as: radius (R), bulk density ($\rho$), escapes velocity ($V_e$) and surface temperature ($T_s$), whereas $x_0$ is with reference to Earth, and $w_x$ is the weight exponent. These parameters are expressed in EU (Earth Units), while the surface temperature is represented as Kelvin (K). 
ESI is basically classified into interior ESI and surface ESI, which are mathematically represented as:
\begin{equation}
ESI_I = \sqrt{{ESI_R} \times {ESI_\rho}}\,,
\label{eq:interiorESI}
\end{equation}
and surface ESI is:
\begin{equation}
ESI_S = \sqrt{ESI_T \times ESI_{V_e}}\,,
\label{eq:surfaceESI}
\end{equation}
where $ESI_R$, $ESI_\rho$, $ESI_T$ \&  $ESI_{V_e}$ are Earth Similarity Indices, calculated for radius, density, surface temperature and escape velocity, respectively. The global ESI is their geometric mean: 
\begin{equation}
ESI = \sqrt{{ESI_I} \times {ESI_S}}\,.
\label{eq:globalESI}
\end{equation}

From the ESI analysis, 44 potentially habitable planets were identified, keeping Mars as the threshold with ESI value 0.73. 

\section{Colonizing Similarity Index (CSI)}
CSI is basically representing Earth-like planets that is suitable for rocky extremophiles that can survive in extreme temperatures (i.e. as hot as desert and cold as frozen lands). Using equations (2) and (3), the upper and lower weight exponents are chosen for radius, density and escape velocity, similar to ESI (Schulze-Makuch et al., 2011). Mckay, 2014, re-defined the surface temperature in the range 258 - 395 K for extremophiles to reproduce and grow. The CSI average weight exponents are calculated for rocky exoplanets in Table 1. 

\begin{table}[h!]
\begin{center}
\caption{NESI Parametric Table}\label{table:1}
\begin{tabular}{l*{4}{c}r}
\hline
Planetary Property &Ref. Value & Weight Exponents\\
            & for CSI  & for CSI & \\ \hline
Mean Radius	& 1EU  & 0.57 \\ 
Bulk Density &	1EU  &	1.07 \\ 
Escape Velocity	&1EU&	0.70\\ 
Surface Temperature	& 288K& 2.26\\
\hline
\end{tabular}
\end{center}
\end{table}
Here, EU = Earth Units, where Earth’s radius is 6371 km, density is 5.51 g/cm$^3$, escape velocity is 11.19 km/s, Revolution is 365.25 days and surface gravity is 9.8 m/s$^2$.
The main finding in this work is that the weight exponent value of surface temperature is 2.26 for rocky extremophiles to survive, which is found to be lesser than ESI weight exponent value of 5.58 for complex life form to survive (Schulze-Makuch et al., 2011).
CSI is defined for the rock dependent extremophiles, as the geometrical mean of radius, density, escape velocity and surface temperature of exoplanets, with probabilistic range 0 to 1, where “0” indicates non-survival and “1” represents survival. 

Mathematically, the CSI of each physical parameter is defined similar to ESI, which is
\begin{equation}
CSI_x = {\left[1-\Big|
\frac{x-x_0}{x+x_0}\Big| \right]^{w_x}}\,,
\label{eq:esi}
\end{equation}
where, x is the physical parameter of the exoplanet, such as: radius (R), bulk density ($\rho$), escapes velocity ($V_e$) and surface temperature ($T_s$), whereas $x_0$ is with reference to Earth, and $w_x$ is the weight exponent. These parameters are expressed in EU (Earth Units), while the surface temperature is represented as Kelvin (K). 
CSI is basically classified into interior CSI and surface CSI, which are mathematically represented as:
\begin{equation}
CSI_I = \sqrt{{CSI_R} \times {CSI_\rho}}\,,
\label{eq:interiorESI}
\end{equation}
and surface CSI is:
\begin{equation}
CSI_S = \sqrt{CSI_T \times CSI_{V_e}}\,,
\label{eq:surfaceESI}
\end{equation}
where $CSI_R$, $CSI_\rho$, $CSI_T$ \&  $CSI_{V_e}$ are colonizing Similarity Indices, calculated for radius, density, surface temperature and escape velocity, respectively. The global CSI is their geometric mean: 
\begin{equation}
CSI = \sqrt{{CSI_I} \times {CSI_S}}\,.
\label{eq:globalESI}
\end{equation}

The data of radius, density, escape velocity, and surface temperature from Kashyap et al. 2017, is plugged in to equations 9, 10, 11, 12 along with the weight exponent value of surface temperature, 2.26 (present finding), the CSI values are computed (Table 2). 

\begin{table}[h!]
\centering
\caption{A sample of calculated CSI}
\begin{tabular}{l*{8}{c}r}
\hline
Names & Radius & Density &Temp & E. Vel  &$CSI_S$ & $CSI_I$ & CSI\\ 
\ &  (EU) & (EU) & (K) & (EU) &  \ & \ &  \\\hline 
Earth	&1.00	&1.00	&288	&1.00&	1.00&	1.00&	1.00\\
Mars	&0.53	&0.73&	240&	0.45&	0.82&	0.76&	0.79\\
Proxima Cen-b&	1.12&	0.90&	267&	1.06&	0.95&	0.94&	0.95\\
GJ 667Cc	&1.54&	1.05&	288	&1.57&	0.92&	0.92&	0.92\\
Kepler-296 e&	1.48&	1.03&	312&	1.50&	0.93&	0.88&	0.90\\
\hline
\end{tabular}
\label{Table:2}
\end{table}

The entire data table is cataloged and made available online at \citep{Madhu1}.\\

\subsection*{Graphical Representation of rocky planets using Colonizing Similarity Index (CSI)} 

\begin{figure}[h!]
\centering        
\includegraphics[width=10cm,angle=0]{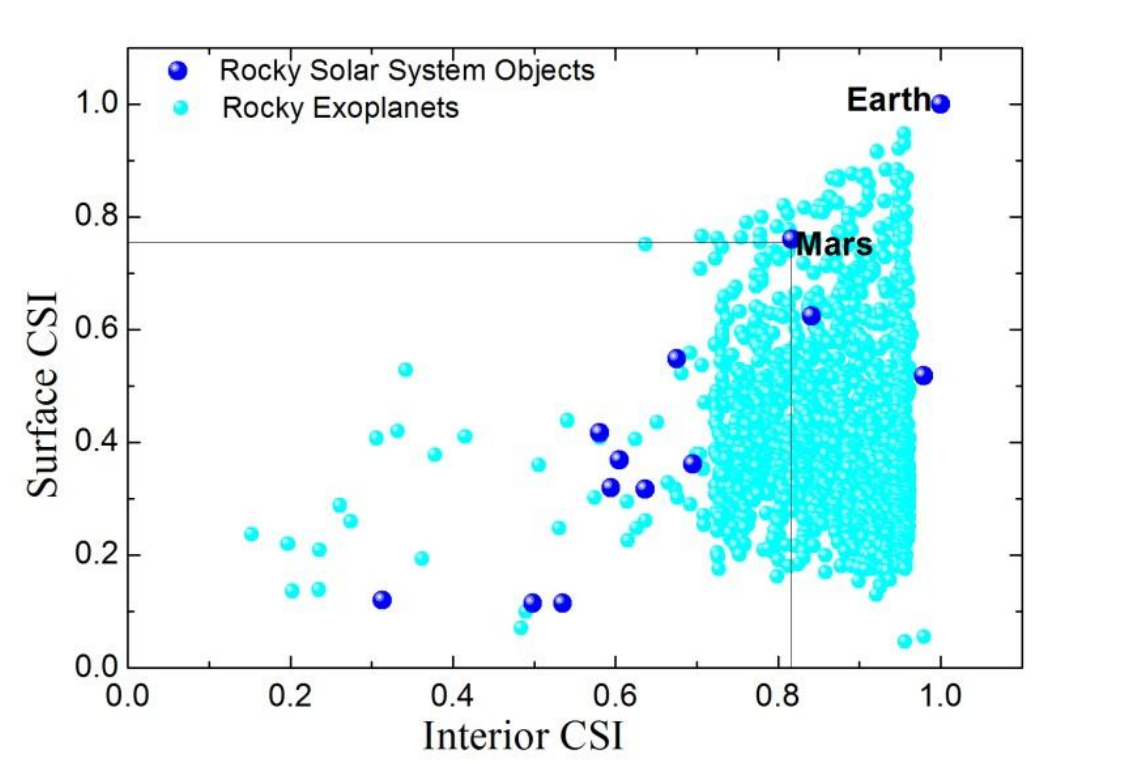} 
\caption{Scatter plot between interior CSI and surface CSI: The solid line marked for Mars will be the threshold value for potentially habitable planets}
\label{fig:CSIplot} 
\end{figure}

The CSI threshold is Mars (~ 0.79), above which the rocky exoplanets are considered to be potentially habitable for extremophiles (such as: Chroococcidiopsis and Acarosporamto) to reproduce and grow.

\newpage 

\section{Discussion and Conclusion}
Earth is the only known rocky planet filled with life, and has been shielded by magnetic field to protect from harmful cosmic radiation (Evans, 1942). In order to search life else-where, we have chosen the rocky exoplanets with reference to Earth conditions and the subjects (Such as: Chroococcidiopsis and Acarosporamto) are targeted to study the reproduction and growth with its high tolerance to resist harmful cosmic radiations in the absence of planet’s magnetic field. Chroococcidiopsis has been selected for the colonizing test on Mars (Expose Mission by Russia) based on its following qualities: can grow on rocks, can produce oxygen, and have tolerance to survive the high energy cosmic radiations (Billi et al. 2000). Acarosporamto is one of the extremophile that has been tested by expose mission for 1.5 years and it has survived the test in space for Mars-like conditions (Onofri et al., 2015).  
According to Kashyap et al. 2017, Mars with ESI value 0.73 was defined as the threshold number for planets to be potentially habitable for complex life forms to survive, whose temperature range was defined from 273 K to 323 K. Approximately 44 potentially habitable planets (php’s) was found till date.
From the CSI formulation, with the threshold 0.79, it is found that 1552 exoplanets are totally in-habitable with the temperature rage in particular for these extremophiles to survive and 98 could be Php’s for extremophiles to survive (provided the other atmospheric conditions are habitable). Presently many works are being performed to prove that Chroococcidiopsis is one of the major candidates for making the red planet green\footnote{$ https://science.nasa.gov/science-news/science-at-nasa/2001/ast26jan_1/$} and Russian Expose mission backs up the candidate choices used in this research. Future missions like James-Webber Space Telescope will give us deeper insights to analyze and work on these potentially habitable planets.

\section*{Acknowledgments} 
I would like to thank Dr. Margarita Safanova former visiting scientist from Indian Institute of Astrophysics (IIA) for all the fruitful discussions. Also a special thanks to Mrs. Vani former (M. Tech Biotechnology) project intern from National Chemical Laboratory (NCL) Pune, for the brief discussions on extremophiles. This research work had made use of PHL-EC, at UPR Arecibo, 2017 {\tt http://phl.upr.edu/projects/habitable-exoplanets-catalogue/data/database}, NASA Exoplanet Archive, and NASA Astrophysics Data System Abstract Service. This work was presented as a part of 105th Indian Science Congress for young scientist award program.
\end{document}